\documentclass[11pt]{article}
\usepackage{moriond,epsfig}
\usepackage{multicol}

\def\be{\begin{equation}}
\def\ee{\end{equation}}
\def\bea{\begin{eqnarray}}
\def\eea{\end{eqnarray}}

\begin{document}
\vspace*{4cm}
\title{HIGGS SEARCHES AND EXTRA DIMENSIONS}

\author{ M. QUIROS }

\address{Instituci\`o Catalana de Recerca i Estudis Avan\c{c}ats (ICREA), Barcelona, Catalonia, Spain\\and\\
Institut de Fisica d'Altes Energies (IFAE), Universidad Aut\`onoma de Barcelona\\ 08193 Bellaterra, Barcelona, Spain}

\maketitle\abstracts{
We present a short review of theories based on warped extra dimensions (motivated by the hierarchy problem of the Standard Model) which can accomodate a Higgs boson in the range suggested by the recent LHC results at 7 TeV. Using the AdS/CFT correspondence the Higgs is composite and can be described in the dual theory by a bound state of the 4D CFT. We have classified the theories in those with a scalar Higgs (5D SM) and those where the Higgs is the fifth component (gauge-Higgs unification) of a bulk gauge field.}

\section{Introduction}
The Large Hadron Collider (LHC) is trying to answer the most fundamental question in particle physics: what is the nature of the electroweak symmetry breaking? Is it a perturbative mechanism, as the Brout-Englert-Higgs (BEH) mechanism, or a non-perturbative one, as in QCD? From the experimental point of view everything seems to be consistent with the Standard Model (SM) with the BEH mechanism (also known as Higgs mechanism) and possibly with a Higgs scalar around 124-126 GeV. However from the theoretical point of view such electroweak vacuum is not stable under quantum corrections (also known as hierarchy problem) which provide
\begin{equation}
\Delta m_H^2=-\frac{3h_t^2}{8\pi^2}\Lambda^2
\end{equation}
where $\Lambda$ is the SM cutoff. In the absence of any tuning this implies an upper bound on the cutoff scale as
\begin{equation}
\Lambda<600\textrm{ GeV}\frac{m_H}{200\textrm{ GeV}}
\end{equation}
and therefore for a larger cutoff there should exist new physics to stabilize it. In fact uncover the nature of the electroweak symmetry breaking should amount to uncovering the kind of new physics (if any) which stabilizes the electroweak vacuum!

There are two main avenues for solving the hierarchy problem:

\begin{description}
\item{\textit{\underline{Elementary Higgs}}}

In this case there should exist an extra symmetry, and new particles with couplings dictated by this symmetry, such that the quadratic sensitivity to high scale cancels. The typical and paradigmatic example is supersymmetry~: the stops cancel the quadratic divergence generated by the top quark. This solution has been covered at this Conference by G.~Altarelli's talk~\cite{Altarelli} and we will concentrate ourselves in the next alternative solution.

\item{\textit{\underline{Composite Higgs}}}

The hierarchy problem is also solved provided that at some scale the Higgs dissolves, and the theory of its constituents is at work. This case is similar to QCD where the pions dissolve into quarks and gluons beyond $\Lambda_{QCD}$. In fact the compositeness scale acts as a cutoff of quadratic divergences. The typical example of compositeness is technicolor. Modern theories of compositeness involve an extra dimension through the
Anti-de-Sitter/Conformal-Field-Theory (AdS/CFT) correspondence~\cite{Maldacena:1997re}.
\end{description}

\section{Extra dimensions and Composite Higgs}
The original AdS/CFT correspondence in string theory~\cite{Maldacena:1997re} related
type IIB string theory on $AdS_5\times S^5$ with $\mathcal N=4\ SU(N)$ 4D gauge theory, the parameters of the correspondence being
\begin{equation}
\left(\frac{M_s}{k}\right)^4=4\pi g^2 N
\end{equation}
where $M_s$ is the string scale, $k$ the AdS curvature and $g$ the gauge coupling constant  of the $\mathcal N=4$ $SU(N)$ supersymmetric theory which is known to be a CFT. In the regime where we can decouple the string excitations and describe the theory by pure gravity $k\ll M_s$ it turns out that $g^2N\gg 1$ which implies that the 4D field theory is non-perturbative. Moreover if the $S^5$ radius is small enough we can decouple its heavy modes and the gravity theory corresponds just to $AdS_5$. 

In the case of a slice of AdS [with two branes, one in the ultraviolet (UV) $y=0$ and another one in the infrared (IR) $y=y_1$ region] a similar correspondence can also be formulated:
The UV boundary corresponds to a UV cutoff in the 4D CFT.
The IR boundary corresponds to an IR cutoff.
Matter localized towards the UV boundary is mainly elementary: e.g. light fermions.
Matter localized towards the IR boundary is mainly composite: e.g. heavy fermions, Higgs boson and Kaluza-Klein (KK) excitations.
%
Although the CFT picture is useful for understanding some qualitative aspects of the theory it is useless for obtaining quantitative predictions since the theory is strongly coupled.

An AdS 5D theory with two branes was proposed long ago~\cite{Randall:1999ee}. In order to solve the hierarchy problem the Higgs should be either:
\textbf{i)} Localized on the IR brane (i.e.~composite).  In that case the theory is disfavored by electroweak precision tests (EWPT);
\textbf{ii)} Propagating in the bulk of the fifth dimension but with a profile leaning towards the IR brane (i.e.~with a certain degree of compositeness).
%
In all cases the hierarchy problem is solved because the Planckian Higgs mass is warped down to the weak scale by the geometry.

A Higgs propagating in the bulk can be:
\begin{description}
\item{\textit{\underline{Scalar-Higgs: H}}}

In this case EWPT require either:
\begin{itemize}
\item
An extra (custodial) gauge symmetry in the bulk generating \textit{non-minimal} models~\cite{Agashe:2003zs}.
\item
A deformation of the AdS metric in the IR~\cite{Cabrer:2009we}. In this case we can consider the \textit{minimal} 5D SM propagating in the bulk.
\end{itemize}
Here we will only consider the latter case as the former one is already covered by the (next) gauge-Higgs unification case.
\item{\textit{\underline{Gauge-Higgs: $A_5$}}}

In this case the Higgs is the fifth component of the gauge boson of an extended gauge group which can possibly contain a custodial symmetry. Now the Higgs bulk mass is doubly protected by the warp factor and by gauge invariance.
\end{description}

\section{Scalar Higgs}
\label{scalar}
The first and simplest possibility is to consider a SM-like Higgs propagating in the 5D space 
\begin{equation}
H(x,y)=\frac{1}{\sqrt 2}e^{i \chi(x,y)} \left(\begin{array}{c}0\\h(y)+\xi(x,y)
\end{array}\right),\quad h(y)=h(0)e^{a ky}
\end{equation}
with an arbitrary metric $A\equiv A(ky)$. The parameters of the effective Lagrangian for the Higgs boson,
\begin{equation}
\mathcal L_{\rm eff}=-|D_\nu   \mathcal H|^2+\mu^2 |\mathcal H|^2-\lambda |\mathcal H|^4
\end{equation}
are related to 5D quantities as
\begin{equation}
\lambda\sim Z^{-2} \,;\quad
\rho= k e^{-A(y_1)};\quad m_H^2
=\frac{2} {Z}\left(M_1/k-a \right)\rho^2\,;\quad
Z=
k\int_0^{y_1}dy\frac{h^2(y)}{h^2(y_1)}e^{-2A(y)+2A(y_1)}
\label{relaciones}
\end{equation}
where $\rho$ is the warped down AdS curvature scale ($\sim$ TeV) and $Z$ is a wave-function renormalization of the Higgs field which appears from integration over the extra dimension. We can see from (\ref{relaciones}) that the natural value of the Higgs mass is $\rho$, so that a light Higgs requires a certain amount of fine-tuning, unless the factor $Z\gg 1$ as we will see it happens in deformed models.

\subsection{RS model}
In the RS model~\cite{Randall:1999ee} the metric is conformally symmetric $A(y) = ky$,
$Z = \mathcal O(1)$ and the natural value of the Higgs mass is TeV. Moreover confronting the model with EWPT implies heavy KK modes (unless extra custodial gauge symmetry in the bulk) and a little fine-tuning problem.
%
\begin{figure}[thb]
\includegraphics[width=8cm,height=6cm]{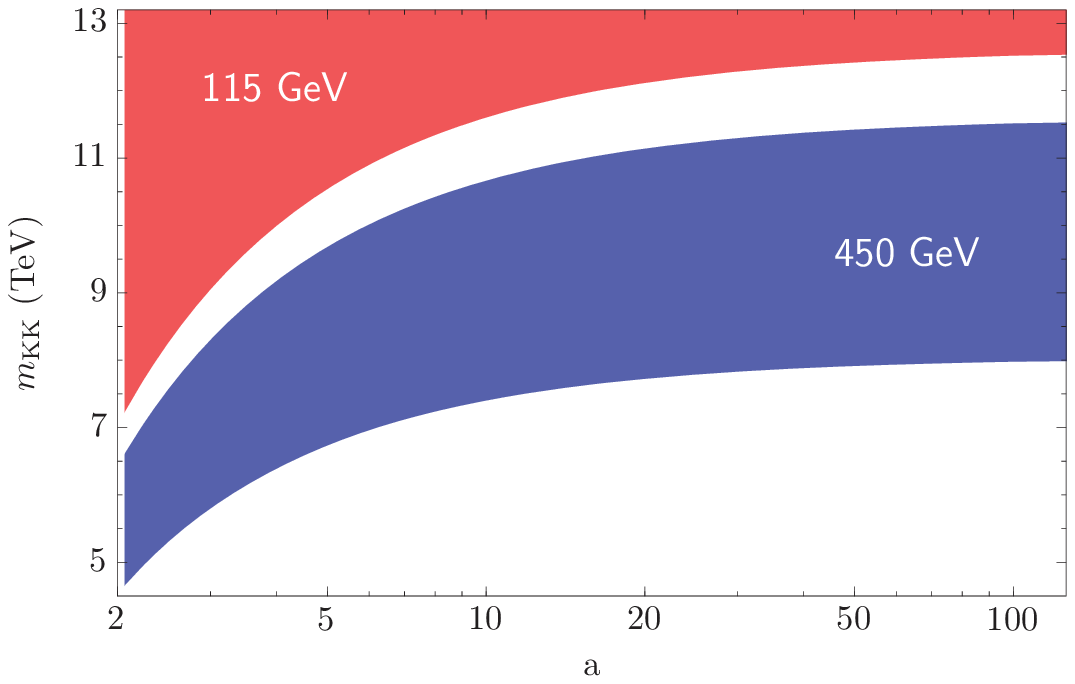}
\includegraphics[width=8cm,height=6cm]{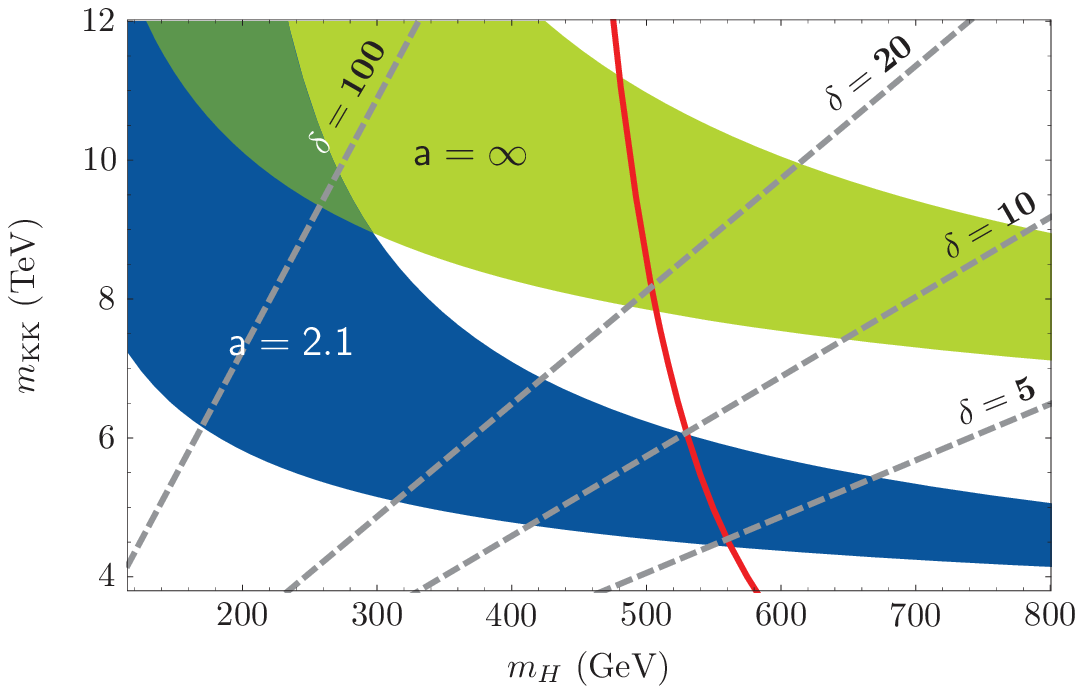}
\caption{Left panel:  Constraints imposed by EWPT on the mass of the first KK mode as a function of $a$ for different values of the Higgs mass. Right panel: The same as a function of $m_H$ for different values of the parameter $a$. We also exhibit the contour lines of constant $\delta$, the sensitivity, defined such that the fine-tuning is $100/\delta\ \%$. }
\label{fig1}
\end{figure}
%
As we can see from Fig.~\ref{fig1} for a Higgs mass $m_H=125$ GeV: \textbf{i)} There is no hope to detect KK modes at LHC since they are too heavy; \textbf{ii)} The fine tuning is at the level of a few per mille.

\subsection{Deformed metric model}

A possible solution to the previous problem is deforming the metric~\cite{Cabrer:2009we}, as in the soft-walls used in AdS/QCD theories. In particular we will consider the metric
\begin{equation}
A(y)=ky-\frac{1}{\nu^2}\log\left(1-y/y_s  \right)
\label{deformed}
\end{equation}
where $\nu$ is a real ($\nu>0$) parameter and which has a singularity at $y=y_s>y_1$, outside the physical interval. One recovers the AdS metric in the limit $\nu\to\infty$ and/or $y_s\to\infty$. Notice also that AdS and the deformed metric (\ref{deformed}) differ only in the IR region while in the UV the metric behaves as AdS and the main features of AdS/CFT duality hold.

In the deformed metric theory the warping is more efficient and consequently the compactification volume is smaller than in the RS theory. This helps in reducing the electroweak precision observables, in particular the oblique observable $T$ which is volume enhanced. Moreover the wave function renormalization parameter is $Z\gg 1$ which helps in: \textbf{i)} Having a light Higgs mass without any fine-tuning, as can be seen from (\ref{relaciones}); \textbf{ii)} Reducing the observables $T$ (which scales as $1/Z^2$) and $S$ (which scales as $1/Z$).
\begin{figure}[thb]
\includegraphics[width=8cm,height=6.2cm]{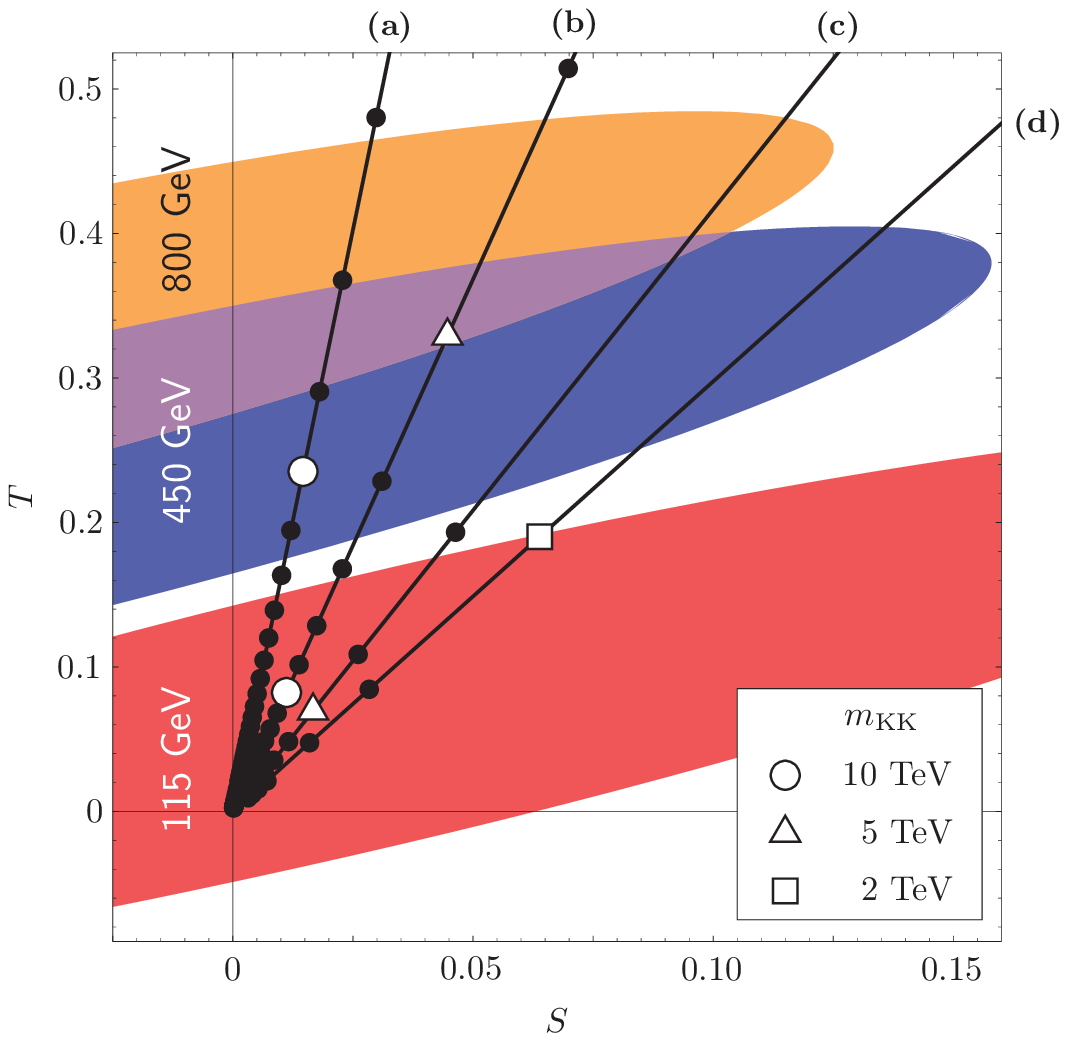}
\vspace{.1cm}\includegraphics[width=8cm,height=6.cm]{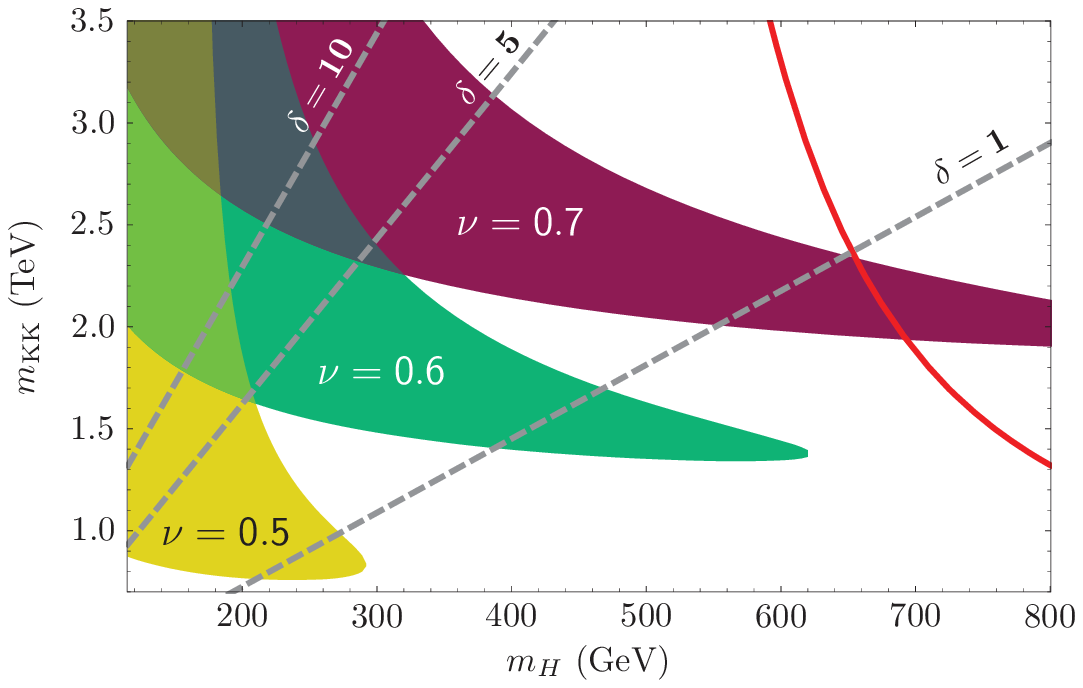}
\caption{Left panel: Constraints imposed by EWPT on the mass of the first KK mode in the plane $(S,T)$ for different values of the Higgs mass. $\Delta m_{KK}=1$ TeV. Right panel:  Constraints imposed by EWPT on the mass of the first KK mode as a function of $m_H$ for different values of the parameter $a$ and contour lines of constant $\delta$.}
\label{fig2}
\end{figure}
As we can see from Fig.~\ref{fig2} for $m_H=125$ GeV the required fine-tuning is better than 10\% and KK-modes have masses of a few TeV so that they can in principle be detected at the LHC. Production of KK modes in this scenario has been analyzed in Refs.~\cite{Carmona:2011ib} which focus on signals
involving the third generation and study in particular $\bar tt$ production, which is dominated by the KK
gluon exchange, as well as $\bar tb$ in the case of the charged KK gauge bosons. After making
adequate cuts, the LHC should be able to probe the existence of the KK gluons for
$\sqrt{s} = 8$ TeV and integrated luminosity of 10$fb^{-1}$ while testing the charged electroweak KK gauge bosons would
require $\sqrt{s }= 14$ TeV and larger luminosities.

To analyze perturbative unitarization of the theory one can compute the coupling of the Higgs to gauge bosons and in
particular one can prove that~\cite{Cabrer:2009we}
\begin{equation}
h_{WWH}^2=h_{WWH,SM}^2\left(1-\xi\right),\quad \xi=\mathcal O(m_H^2/m_{KK}^2)\simeq 0.01
\label{acoplo}
\end{equation}
so a light Higgs unitarizes the theory in a similar way as the SM Higgs.

\section{Gauge-Higgs}
Gauge-Higgs unification is another alternative to supersymmetry where the gauge symmetry in the bulk $\mathcal G$ protects the mass of extra-dimensional components of gauge bosons. This solution to the hierarchy problem requires: \textbf{i)} An extended space-time, for instance a 5D space. \textbf{ii)} An extended gauge group with respect to the SM $SU(3)\otimes SU(2)_L\otimes U(1)_Y$ group. It can be constructed in flat or warped space, although in warped space the GIM-RS mechanism protects the theory with differently localized fermion fields from huge flavor violation, which otherwise would require severe constraints on the mass of KK modes~\cite{Delgado:1999sv}. Four dimensional components of gauge bosons $(A_\mu^a)$ of $\mathcal G$ contain the four-dimensional gauge bosons while the fifth components $(A_5^{\hat a})$ contain the four-dimensional Higgs fields in a number equal to the number of Pseudo Goldstone Bosons (PGB) which are left out in the four dimensional theory. In general $\mathcal G$ will be broken by boundary conditions to $\mathcal H_{UV}$ ($\mathcal H_{IR}$) on the UV (IR) brane. For $\mathcal H_{UV}=SU(2)_L\otimes U(1)_Y$ the number of PGB is $\dim(\mathcal G/\mathcal H_{IR})$ so different models differ by different choices for $\mathcal G$ and $\mathcal H_{IR}$. Some models~\cite{Contino:2003ve} are defined in the table below.
\vspace{.5cm} 
  \begin{center}
\begin{tabular}{||c|c||}
\hline\noalign{\smallskip}
Model & \# Goldstones ($A_5^{\hat a}$)  \\
\hline
{SO(4)}/{SO(3)} & 6-3=3 (Higgsless SM)\\
{SU(3)}/{SU(2)$\times$U(1)} & 8-4=4$(H_{SM})$\\ 
{SO(5)}/{SO(4)} & 10-6=4 $(H_{SM})$\\
{SO(6)}/{SO(5)} & 15-10=5 ($H_{SM}$ + singlet)\\
{SO(6)}/{SO(4)$\times$SO(2)} & 15-6-1=8 $(H_u,H_d)$\\
\noalign{\smallskip}\hline
\end{tabular}
\end{center}
%
\vspace{.5cm} Some of the models in the table contain the custodial $SO(4)$ group on the IR brane and so their contribution to the $T$ parameter is protected. In these theories $SU(2)_L\otimes U(1)_Y$ breaking is radiative
through the kind of diagrams~\cite{vonGersdorff:2002as} exhibited in Fig.~3.
\begin{figure}[thb]
\begin{center}
\includegraphics[width=0.75\textwidth]{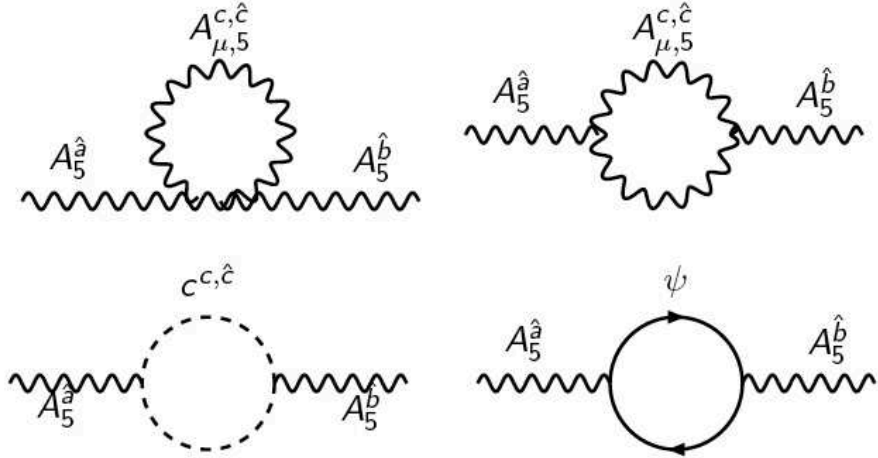}
\caption{One-loop diagrams contributing to electroweak breaking in gauge-Higgs unification models. }
\end{center}
\label{fig3}
\end{figure}
It turns out that triggering electroweak breaking will depend on the nature and localization of bulk matter fields.

In the dual theory $\mathcal G/\mathcal H_{IR}$ is characterized by the spontaneous breaking scale $f$ such that the expansion parameter in the theory is $\xi$
\begin{equation}
\xi\equiv\left(\frac{v}{f}\right)^2\quad \left\{ \begin{array}{l} \bullet\quad \xi\to 0 \Rightarrow \textrm{SM limit} \\\bullet\quad \xi\to 1 \Rightarrow \textrm{Technicolor limit}\end{array}\right.
\end{equation}
where $v\simeq 246$ GeV is the electroweak breaking parameter. The $\xi$ parameter controls perturbative unitarity through the relation (\ref{acoplo}). However unlike in the models presented in Sec.~\ref{scalar} with a scalar Higgs, where the parameter $\xi\ll 1$, in the models presented in this section $\xi$ depends on $f$ and can thus be considered as a free parameter. For instance in the limit $\xi\to 0$ the SM result is obtained and the Higgs unitarizes the theory without the need of any extra particle. On the other extreme in the Technicolor limit $\xi\to 1$ all unitarity must be provided by new TeV resonances at scales close to the electroweak scale. For intermediate values of $0<\xi<1$ unitarity must be partially restored by resonances at scales which depend on the value of $\xi$. 

One can consider in general an effective theory~\cite{Espinosa:2010vn} parametrized by $\xi$, which measures the degree of compositeness of the Higgs. In this theory all Higgs couplings (cubic, quartic, $HWW$, \dots) depart from the SM values by quantities which are proportional to $\xi$. These models have been confronted to EWPT and direct searches~\cite{Espinosa:2012qj}. The former ones provide the strongest constraints which yield typical bounds $\xi< 0.18$ at 99\% CL in the absence of additional contributions to the $S$ and $T$ parameters.

\section{Conclusions}
It is clear at the moment of this Conference that the next step in Higgs search belongs to the LHC Collaborations, in particular ATLAS and CMS, so that confronting different theories on electroweak breaking with experimental data should wait till the excess of events found at 125 GeV, which hints on the presence of a Higgs boson, be eventually confirmed. We can only speculate about the different possibilities. If Higgs is confirmed at $m_H\simeq 125 $ GeV and cross sections in different channels are consistent with the SM expectations, then the SM is a good candidate and there should be no problem with perturbative unitarity or (meta)stability of the electroweak vacuum.  If Higgs is confirmed at $m_H\simeq 125 $ GeV and cross sections in some channels are \textit{not} consistent with the SM expectations (as it seems to be the case now with $\gamma\gamma_{ggF}$ excess) then one should consider to extend the SM to theories with a light Higgs and extra matter which can eventually modify some of the relevant production cross sections (supersymmetry, composite Higgs, extra dimensions,  unexpected physics, \dots). Finally if it turns out that the Higgs is not found then it might be heavy in which case extra states should soon appear.

\section*{Acknowledgments}
Work supported in part by the Spanish Consolider-Ingenio 2010 Programme CPAN (CSD2007-00042) and by CICYTFEDER-FPA2008-01430 and FPA2011-25948.

\end{document}